\documentclass[prl,twocolumn,showpacs]{revtex4}
\pdfoutput=1
\usepackage{amsmath}
\usepackage{amssymb}
\usepackage{graphicx}

\newcommand{\ket}[1]{\mbox{$\vert #1 \rangle$}}
\newcommand{\bra}[1]{\mbox{$\langle #1 \vert$}}

\begin{document}
\title{Microwave Control of Atomic Motion in Optical Lattices}
    \author{Leonid F\"orster}
    \author{Micha\l\ Karski}
    \author{Jai-Min Choi}
    \author{Andreas Steffen}
    \author{Wolfgang Alt}
    \author{Dieter Meschede}
    \author{Artur Widera} \email{widera@uni-bonn.de}
    \address{Institut f\"ur Angewandte Physik, Universit\"at Bonn,
    Wegelerstr.~8, D-53115 Bonn, Germany}
    \author{Enrique Montano}
    \author{Jae Hoon Lee}
    \author{Worawarong Rakreungdet}
    \author{Poul S. Jessen}
    \address{CQuIC, College of Optical Sciences, University of Arizona, Tucson, Arizona 85721, USA}
%k   
\date{\today}
\pacs{37.10.De, 37.10.Jk, 37.10.Vz, 05.60.Gg}

\begin{abstract}
We control the quantum mechanical motion of neutral atoms in an optical lattice by driving microwave transitions between spin states whose trapping potentials are spatially offset. Control of this offset with nanometer precision allows for adjustment of the coupling strength between different motional states, analogous to an adjustable effective Lamb-Dicke factor. This is used both for efficient one-dimensional sideband cooling of individual atoms to a vibrational ground state population of 97\%, and to drive coherent Rabi oscillation between arbitrary pairs of vibrational states. We further show that microwaves can drive well resolved transitions between motional states in maximally offset, shallow lattices, and thus in principle allow for coherent control of long range quantum transport.
\end{abstract}

\maketitle

Accurate, simultaneous control of multiple degrees of freedom is crucial for the experimental realization of quantum information processing and quantum simulation \cite{nielsen_quantum_2000}.   Thus, in schemes that use trapped ions or atoms as carriers of quantum information, qubits are often encoded in internal states and the motional degree of freedom is manipulated to engineer state dependent interactions between them.  In ion traps this involves the controlled excitation of a collective mode of vibration \cite{cirac_quantum_1995,monroe_demonstration_1995}, leading to robust quantum gate protocols and entanglement of multiple qubits \cite{leibfried_creation_2005,haffner_scalable_2005}.  Proposals based on neutral atoms in optical lattices rely instead on short range collisional interactions \cite{jaksch_entanglement_1999,brennen_quantum_1999}, which can be controlled by overlapping the center of mass wavepackets.  Variations of this approach have been used to entangle pairs \cite{anderlini_controlled_2007} and chains of atoms \cite{mandel_controlled_2003}, with a fidelity limited mainly by the speed and accuracy with which the optical potentials can be changed during the transport phase.  For this reason there has been considerable interest in optimal control techniques to improve atom transport in optical lattices and other traps \cite{calarco_quantum_2000,creffield_quantum_2007,merkel_coherent_2007,chiara_optimal_2008}, mostly through more elaborate control of the trapping potentials.  Modulation of optical lattice potentials has also been explored as a way to control tunneling \cite{lignier_dynamical_2007,kierig_single-particle_2008}.  
Here we show that controlling the relative position of state-dependent lattices allows to manipulate the motion of atomic qubits in static potentials \cite{deb_method_2007} using microwaves. In principle this enables the generation of a broad class of spin-motion entangled states \cite{haroutyunyan_coherent_2001,mischuck_coherent_2009}, and lends itself readily to the application of composite pulses and other robust control techniques (see Ref.~\cite{rakreungdet_accurate_2009} and references therein).
\begin{figure}
          \centering
          \includegraphics [scale=1]{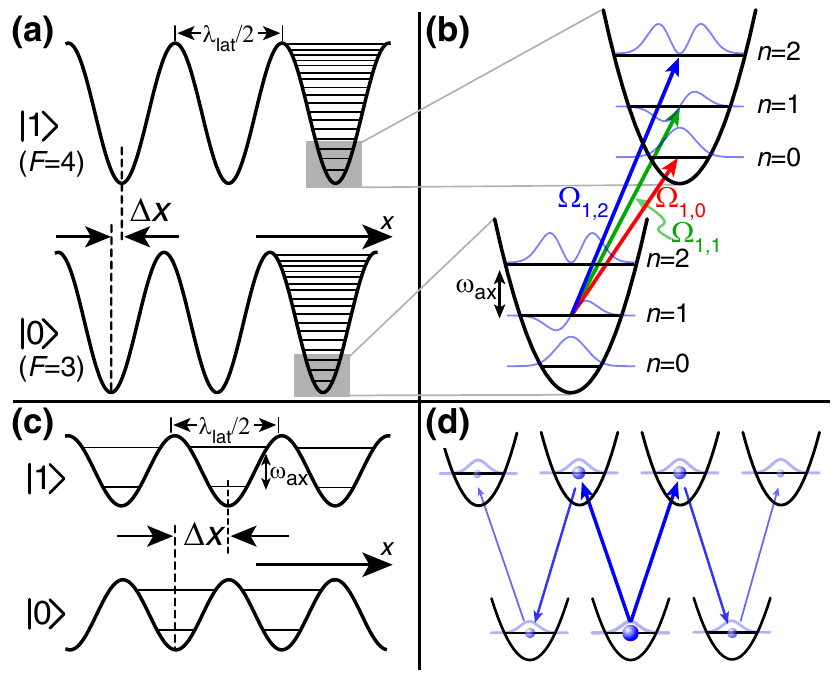}
          \caption{(a) Deep, slightly offset state-dependent optical lattices for the \ket{0} and \ket{1} hyperfine states. (b) The offset enables on-site microwave transitions changing the vibrational quantum number. (c) Maximally offset, shallow lattices. (d) Microwave radiation can couple an initially localized population of \ket{0} symmetrically to the neighboring potential wells and further throughout the lattice.}
          \label{fig:Figure1}
\end{figure}

We consider Caesium (Cs) atoms with two hyperfine states $\ket{0} \equiv \ket{F=3,m_F=3}$ and $\ket{1} \equiv \ket{F=4,m_F=4}$, which are coupled by microwave radiation. Here $F$ and $m_F$ are the total angular momentum and its projection onto the quantization axis, respectively. Microwaves provide a coherent, homogeneous and readily controllable radiation field, but since microwave photons have negligible momentum they are rarely considered for driving transitions between vibrational states in atom and ion traps.  We circumvent this limitation by trapping atoms in the $\ket{0}$ and $\ket{1}$ states in separate optical lattice potentials offset by a distance $\Delta x$.  The matrix element for a vibrational transition is then proportional to the center-of-mass wave function overlap, i.e.~the Franck-Condon factor,
\begin{equation}\label{eq:TransitionElement}
	\hbar\Omega_{n,n^\prime} = \hbar\Omega_0\bra{\tilde{n}^\prime} e^{-i \Delta x\, \hat{p}/\hbar} \ket{\tilde{n}} = \hbar\Omega_0\bra{n^\prime}n\rangle,
\end{equation}
where $\Omega_0$ is the "bare" Rabi frequency for the $\ket{0}$ to $\ket{1}$ transition in free space, $n^\prime$ and $n$ ($\tilde{n}^\prime$ and $\tilde{n}$) label the vibrational quantum states in the shifted (unshifted) $\ket{0}$ and $\ket{1}$ potentials, and $\hat{p}$ is the momentum operator. From Eq.~(\ref{eq:TransitionElement})  an effective Lamb-Dicke parameter $\eta_\mathrm{eff}\equiv i \Delta x\,p_0/\hbar$ can be defined, where $p_0$ is the size of the harmonic oscillator wave packet in momentum space. This parameter plays a role analogous to the usual Lamb-Dicke parameter for optical transitions in traps with no spatial offset. Changing the displacement $\Delta x$ changes the effective Lamb-Dicke parameter, and allows us to tune the coupling between an arbitrary pair of vibrational states from zero up to $\sim \hbar\Omega_0/2$.

Comparing the size of the ground state wave packet $x_0$ with the lattice spacing $a_\mathrm{lat}$, one can identify two regimes where the physics is qualitatively different. For deep lattices the atomic wave packets are strongly localized, and the overlap between different vibrational states is only significant for small displacements $\Delta x \ll a_\mathrm{lat}$.
In this situation the atomic motion is confined to a double-potential well consisting of one well of the $\ket{0}$ lattice and one well of the $\ket{1}$ lattice, see Fig.~\ref{fig:Figure1}a,b. This regime is similar to the recently reported coupling of a charge-phase qubit to an \emph{LC} oscillator \cite{gunnarsson_vibronic_2008} or the coupling of motional and spin states of trapped ions using rf radiation and a static magnetic field  \cite{johanning_individual_2009}.

For shallow lattices the extent of the wave packet becomes comparable to half the lattice spacing ($x_0\lesssim a_\mathrm{lat}/2$). In this regime the overlap between the $\ket{0}$ and $\ket{1}$ wave packets is significant even when the lattices are offset by the maximum amount, $\Delta x = a_\mathrm{lat}/2$.  In that case the microwave field does not couple isolated pairs of potential wells, but instead introduces nearest-neighbor coupling between potential wells throughout the $\ket{0}$ and $\ket{1}$ lattices as illustrated in Fig.~\ref{fig:Figure1}c,d. The result is a quantum walk of the atom on the lattice \cite{karski_quantum_2009}, equivalent to ballistic tunneling in a $\lambda/4$ period microwave dressed lattice potential.  

\begin{figure}
          \centering
          \includegraphics [scale=1]{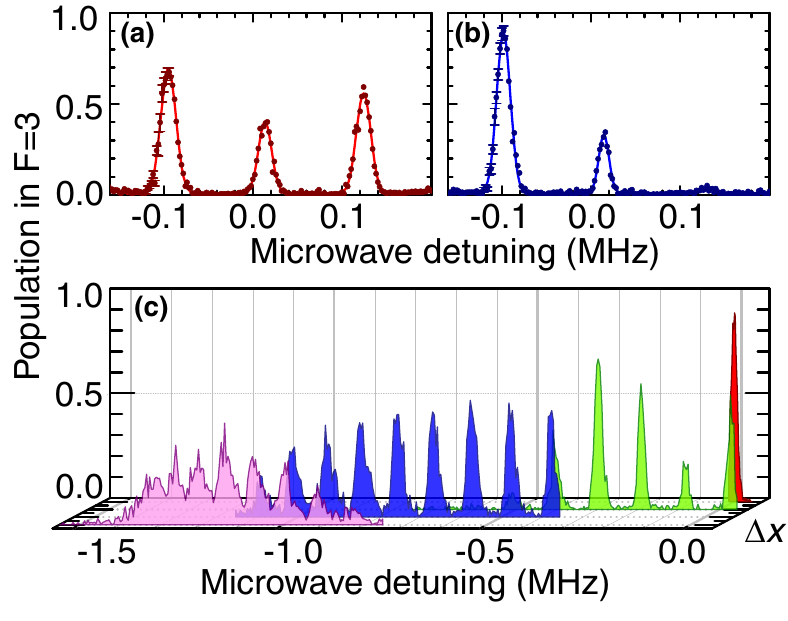}
          \caption{Microwave spectra for strong confining and slightly displaced ($\Delta x = 24\,$nm) traps, for (a) molasses cooled and (b) sideband cooled atoms with a ground state population of 97\%. (c) Spectrum for all vibrational states in the trap, starting from the axial ground state. The different spectra correspond to (from back to front) $\Delta x = 0\,$nm, $\Delta x = 43\,$nm, $\Delta x = 111\,$nm and $\Delta x = 176\,$nm.}
          \label{fig:Figure2}
\end{figure}

We create our lattices by superimposing a pair of counterpropagating laser beams with linear polarizations forming an angle $\vartheta$. The resulting light field consists of two circularly polarized standing waves, which can be shifted in opposite directions along the axis by adjusting the angle $\vartheta$.  Because of the different tensor polarizabilities of the $\ket{0}$ and $\ket{1}$ states, this leads to spatial offset of their respective lattice potentials \cite{mandel_coherent_2003}.  Microwave radiation around 9.2\,GHz couples the $ \ket{0}$ and $\ket{1}$ states with a bare Rabi frequency $\Omega_0$ of up to $60\,$kHz, while a magnetic field of up to 3\,G lifts their degeneracy with other Zeeman states. Spectra are obtained by scanning the frequency of the microwave field and detecting the number of atoms in state \ket{0} or state \ket{1}.  For $\Delta x = 0$, transitions to different vibrational levels are not detectable, but when the displacement is sufficient for the effective Lamb-Dicke parameter $\eta_\mathrm{eff}$ to be non-negligible, the spectrum consists of both a carrier ($\ket{n_\mathrm{ax}} \leftrightarrow \ket{n_\mathrm{ax}}$ transitions) and sidebands at $ \pm \omega_\mathrm{ax}$ ($\ket{n_\mathrm{ax}} \leftrightarrow \ket{n_\mathrm{ax} \pm 1}$ transitions), as seen in Fig.~\ref{fig:Figure2}. 

Strong confinement is realized in a one-dimensional (1D) geometry, in which two counterpropagating laser beams with a wavelength of $\lambda_\mathrm{lat}=865.9$\,nm are tightly focused to a waist of 20\,$\mu$m. After molasses cooling and subsequent adiabatic lowering of the trap depth to $k_B\times 80\,\mu$K, the atoms have a typical temperature of $10\,\mu$K. For our axial and radial trapping frequencies of $\omega_\mathrm{ax} = 2\pi \times 110\,$kHz and $\omega_\mathrm{rad} = 2\pi \times 1.1\,$kHz, respectively, the atoms populate vibrational states with a mean quantum number of $\bar{n}_\mathrm{ax}=1.2$ axially and $\bar{n}_\mathrm{rad} = 200$ radially.

Weak confinement is realized in a shallow, three-dimensional optical lattice consisting of three individual 1D lattices whose optical frequencies differ by tens of MHz.  The lattice is loaded with $10^6$ Cs atoms, which are sideband cooled \cite{hamann_resolved-sideband_1998} into the \ket{0} state, with a mean vibrational excitation of $0.1 - 0.2$ for each dimension. During microwave spectroscopy we set $\vartheta=0$ for the transverse lattices, and the lattice depths are adjusted to obtain vibrational frequencies $\omega_\mathrm{ax}=\omega_\mathrm{trans}=2\pi \times 18\,$kHz. The common lattice detuning is 140\,GHz blue of the Cs $D_2$ line, where the $\ket{0, n=0} \leftrightarrow \ket{1, n=\{0,1\}}$ transition frequencies show minimal dependence on lattice depth (see Fig.~\ref{fig:Figure5}a), and where transition broadening due to lattice beam inhomogeneity is suppressed.

%Different from previous cooling in state dependent traps \cite{miller_rf-induced_2002}, we drive sideband transitions from \ket{1,n_\mathrm{ax}} to \ket{0,n_\mathrm{ax}-1}. 

In the strongly confining lattice we begin our experiments by sideband cooling the atoms with a microwave field tuned to the \ket{1,n_\mathrm{ax}} to \ket{0,n_\mathrm{ax}-1} transitions. This corresponds to the blue sideband of the spectrum as the cooling cycle starts in the upper hyperfine state. Repumping on the $F=3 \rightarrow F^\prime=4$ transition of the $D_2$ line breaks the coherence of the microwave transition, and subsequent spontaneous emission brings the atom back to the \ket{1} ground state. We apply the microwave field and repumping laser simultaneously for 20\,ms \cite{footnote_cooling}. In addition, a second, circularly polarized laser beam on the $F=4 \rightarrow F^\prime=4$ transition ensures spin polarization in state \ket{1}. A spectrum of sideband cooled atoms is shown in Fig.~\ref{fig:Figure2}b.  By comparing the areas of the blue and red sideband peaks \cite{diedrich_laser_1989} we find 97\% axial ground state population. The cooling is limited by the optical transition and emission cycle, which only approximately maintains the vibrational quantum number due to the displaced potentials \cite{footnote_cooling}. This leads to a minimum mean quantum number of $n_\mathrm{ax} = 0.03$.

Vibrational spectra can be obtained by adjusting the coupling strength to favor transitions from the axial ground state to a chosen group of final states.  Fig.~\ref{fig:Figure2}c shows a series of four such spectra measured for different $\Delta x$, mapping out all bound states in our trap.  The experimentally measured transition frequencies are in agreement with a one-dimensional band structure calculation, but the observed line widths are greater than expected from the calculated band widths. We attribute this to inhomogeneous broadening associated with the (uncooled) radial motion of the atoms.
\begin{figure}
          \centering
          \includegraphics [scale=1]{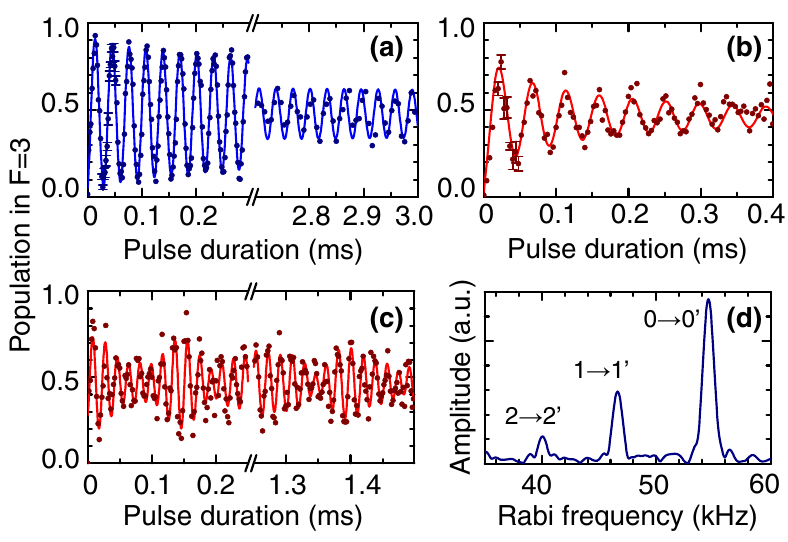}
          \caption{Rabi oscillations for the transitions from \ket{1,0} to (a) \ket{0,1} and (b) \ket{0,7}. The corresponding Rabi frequencies are $2\pi \times \{32,22\}$ kHz, respectively. (c) Rabi oscillations on the carrier for a thermal initial state and (d) its Fourier transform. From the amplitudes we deduce a one-dimensional temperature of $7\,\mu$K. Solid lines are sines damped by an empirically found function to account for radial dynamics \cite{footnote_cooling}.}
          \label{fig:Figure3}
\end{figure}

Coherent Rabi oscillations for the first and the seventh red sideband transitions starting from the $\ket{1,0}$-state are shown in Fig.~\ref{fig:Figure3} a,b. Their Rabi frequencies can be adjusted by the displacement $\Delta x$ to be on the same order of magnitude as the bare Rabi frequency of 60\,kHz.   It is also instructive to observe Rabi oscillations that start from a thermal distribution of vibrational states.  In that case the signal is a sum of Rabi oscillations at a series of distinct frequencies, each corresponding to a transition $\ket{0,n} \leftrightarrow \ket{1,n}$ and contributing in proportion to the population of the initial state.  Fig.~\ref{fig:Figure3}c shows such a beat signal obtained for  $\Delta x= 15\,$nm.  From its Fourier transform (Fig.~\ref{fig:Figure3}d) we deduce a temperature of $7\,\mu$K corresponding to a mean axial vibrational quantum number of $n_\mathrm{ax}=0.8$ \cite{meekhof_generation_1996}.

\begin{figure}
          \centering
          \includegraphics [scale=1]{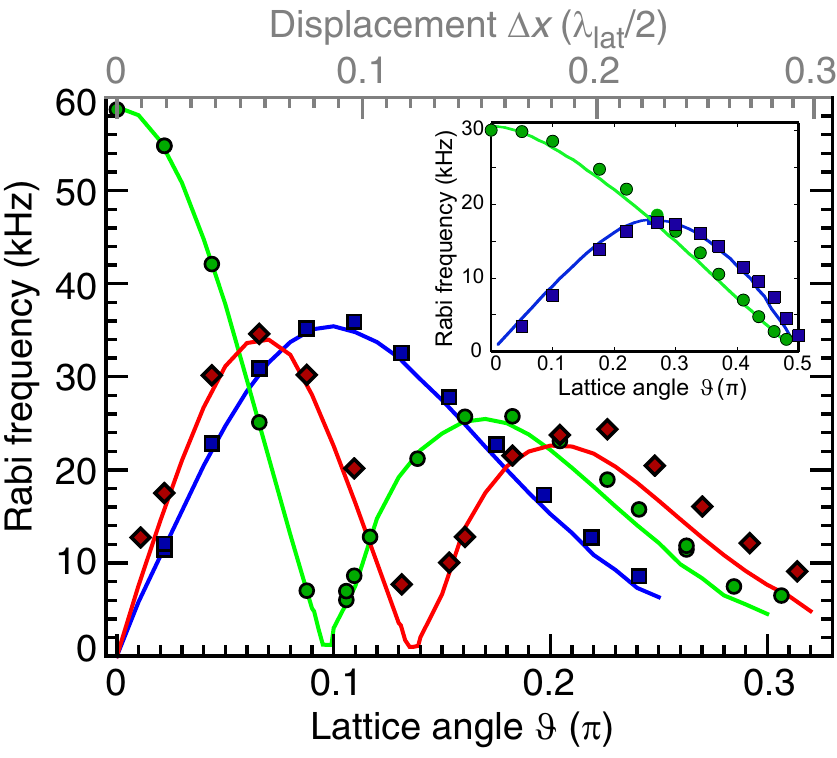}
          \caption{Measured Rabi frequencies of the carrier ($\circ$) and the red ($\diamond$) and blue ({\tiny $\square$}) sidebands, starting from the state $\ket{0,1}$, depending on the polarization angle $\vartheta$. The solid lines are a predictions of a full model. $\Delta x$ depends in a non-linear way on the polarization angle $\vartheta$ and is shown in the upper horizontal axis.  The insert shows measured and calculated Rabi frequencies in a shallow lattice, for angles up to $\vartheta=\pi/2$.}
          \label{fig:Figure4}
\end{figure}

\begin{figure}
          \centering
          \includegraphics[scale=1]{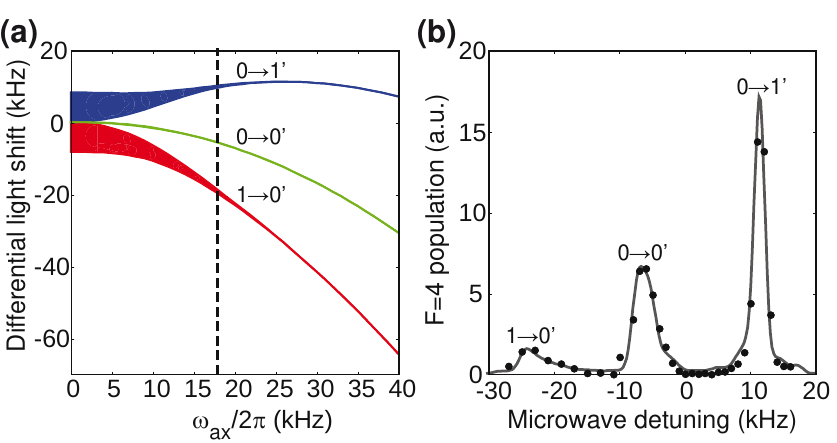}
          \caption{Vibrational spectroscopy in a weakly confining lattice with $\Delta x = \lambda/4$.  (a) Calculated light shift of the $\ket{0, n=0} \leftrightarrow \ket{1, n=\{0,1\}}$ carrier and sideband transitions.  The shaded areas indicate line broadening due to band curvature.  The dashed line indicates the $\omega_\mathrm{ax}$ used in our experiment.  (b) Microwave spectrum.  Solid points are experimental data, the line a prediction from a full model that includes broadening from lattice inhomogeneity.  The various line widths reflect the slope of the light shift curves in (a).}
          \label{fig:Figure5}
\end{figure}

To illustrate the range of adjustable coupling strength, we measure the Rabi frequencies of three transitions starting from the $\ket{0,1}$-state (the carrier and the first red and blue sidebands) for different displacements $\Delta x$, see Fig.~\ref{fig:Figure4}. For weak coupling, ($\Omega_0/\omega_\mathrm{ax} \ll 1$), this yields direct information on the wave function overlap between the initial and final wave functions according to Eq.~(\ref{eq:TransitionElement}). Since our data is not taken in this regime, we compare instead to a theoretical model that takes into account microwave dressing, off-resonance excitation and, for the strong confining regime, radial thermal motion.  This is done by integrating the Schr\"odinger equation in the Bloch basis, and determining the Rabi frequency from the time dependent oscillation of the $\ket{0}$ and $\ket{1}$ states.  For most displacements, our model predictions differ negligibly from Eq.~(\ref{eq:TransitionElement}), and reproduce our experimental data for strong and weak confinement quite well.

In our weakly confining lattice a small but non-zero Rabi frequency remains even for displacements of $\Delta x = \lambda/4$, as shown in the insert in Fig.~\ref{fig:Figure4}. We stress that in a 3D lattice, spatial inhomogeneities of the trapping potential can easily broaden the transition line, but that for the lattice depth and detuning chosen here this sensitivity is suppressed (Fig.~\ref{fig:Figure5}a).  Figure~\ref{fig:Figure5}b  shows an example of a measured spectrum in a lattice with $\Delta x = \lambda/4$, along with a theoretical prediction from our model based on integrating the Schr\"odinger equation in the Bloch basis and taking into account inhomogeneous broadening from variations in the lattice depth and magnetic field across the atomic ensemble.  The excitation pulse has a Gaussian envelope with a full width at half maximum of $1$ ms, and pulse areas of $0.35\pi$ for the $\ket{n_\mathrm{ax}=0} \leftrightarrow \ket{n_\mathrm{ax}=0}$ transition and $0.9\pi$ for the $\ket{n_\mathrm{ax}=0} \leftrightarrow \ket{n_\mathrm{ax}=1}$ transition, respectively.  In this geometry, simultaneous coupling between states throughout the lattice (Fig. \ref{fig:Figure1}c,d) will drive a quantum walk that delocalizes an atom in space on a timescale comparable to the Rabi period.  At this point tracing over the spatial degree of freedom leads to a statistical mixture of the $\ket{0}$ and $\ket{1}$ states.  Experimentally, we see this as a rapid damping of the Rabi oscillation.  The demonstration of clearly resolvable lines in the microwave spectrum is a prerequisite for control of quantum transport, as explored theoretically in \cite{mischuck_coherent_2009}.

In summary, we have demonstrated microwave control of atomic motion in state dependent optical lattices.  We have used this to implement a convenient scheme for sideband cooling, and to drive coherent Rabi oscillations between selected pairs of vibrational states.  In the near term our approach may prove useful in optical lattice based quantum simulation, e.g.~by populating high-lying bands and controlling the tunneling properties in deep lattice potentials.  Possible applications include investigations of non-equilibrium systems \cite{jarzynski_nonequilibrium_1997} on the border between classical and quantum thermodynamics.  Other prospects include detection and control of two-body interactions and trap induced resonances \cite{stock_quantum_2003}.  In the longer term, microwave driven quantum transport may be a good candidate for robust control, which will likely prove essential to quantum information processing.

We acknowledge financial support by DFG (research unit 635), the EU (IP SCALA), the NSF (PHY-0555673) and IARPA (DAAD19-13-R-0011). MK acknowledges support by Studienstiftung des deutschen Volkes, JMC from the Korean Government (MOEHRD).


\begin{thebibliography}{32}
\bibitem{footnote_cooling}
\bibinfo{howpublished}{Details will be presented in an upcoming publication.}
\end{thebibliography}
\end{document}